\documentclass[conference]{IEEEtran}
\IEEEoverridecommandlockouts
\usepackage{cite}
\usepackage{amsmath,amssymb,amsfonts}

\usepackage{etoolbox}

\usepackage{algorithm}
\usepackage{algorithmicx}

\algdef{SE}[DOWHILE]{Do}{doWhile}{\algorithmicdo}[1]{\algorithmicwhile\ #1}%
\usepackage[skip = 1.5pt,labelfont=bf]{caption}

\usepackage[noend]{algpseudocode}
\usepackage{mathtools}

\usepackage[font=small,labelfont=bf]{caption}

\usepackage{graphicx}
\usepackage{textcomp}
\def\BibTeX{{\rm B\kern-.05em{\sc i\kern-.025em b}\kern-.08em
    T\kern-.1667em\lower.7ex\hbox{E}\kern-.125emX}}
\begin{document}

\title{\huge TraNNsformer: \underline{N}eural \underline{N}etwork \underline{Transform}ation for Memristive Crossbar based Neuromorphic System Design \\
\thanks{The work was supported in part by, Center for Spintronic Materials, Interfaces, and Novel Architectures (C-SPIN), a MARCO and DARPA sponsored StarNet center, by the Semiconductor Research Corporation, the National Science Foundation, Intel Corporation and by the Vannevar Bush Faculty Fellowship.}
}

\author{\IEEEauthorblockN{Aayush Ankit, Abhronil Sengupta, Kaushik Roy}
\IEEEauthorblockA{\textit{School of Electrical and Computer Engineering, Purdue University} \\
\{aankit, asengup, kaushik\}@purdue.edu}
}
\maketitle

\begin{abstract}
Implementation of Neuromorphic Systems using post Complementary Metal-Oxide-Semiconductor (CMOS) technology based Memristive Crossbar Array (MCA) has emerged as a promising solution to enable low-power acceleration of neural networks. However, the recent trend to design Deep Neural Networks (DNNs) for achieving human-like cognitive abilities poses significant challenges towards the scalable design of neuromorphic systems (due to the increase in computation/storage demands). Network pruning \cite{han2015learning} is a powerful technique to remove redundant connections for designing optimally connected (maximally sparse) DNNs. However, such pruning techniques induce irregular connections that are incoherent to the crossbar structure. Eventually they produce DNNs with highly inefficient hardware realizations (in terms of area and energy). In this work, we propose TraNNsformer - an integrated training framework that transforms DNNs to enable their efficient realization on MCA-based systems. TraNNsformer first prunes the connectivity matrix while forming clusters with the remaining connections. Subsequently, it retrains the network to fine tune the connections and reinforce the clusters. This is done iteratively to transform the original connectivity into an optimally pruned and maximally clustered mapping. We evaluated the proposed framework by transforming different Multi-Layer Perceptron (MLP) based Spiking Neural Networks (SNNs) on a wide range of datasets (MNIST, SVHN and CIFAR10) and executing them on MCA-based systems to analyze the area and energy benefits. Without accuracy loss, TraNNsformer reduces the area (energy) consumption by 28\% - 55\% (49\% - 67\%) with respect to the original network. Compared to network pruning, TraNNsformer achieves 28\% - 49\% (15\% - 29\%) area (energy) savings. Furthermore, TraNNsformer is a technology-aware framework that allows mapping a given DNN to any MCA size permissible by the memristive technology for reliable operations.
\end{abstract}


\section{Introduction and Related Work}
The quest to reverse-engineer the powerful cognitive capabilities of the human brain inspired the research in artificial neural networks. Although, the precise structure and functioning of the human brain has not been clearly understood, there is sufficient evidence that suggests the hierarchical organization of neurons (cells) in the brain \cite{diehl2015fast}. Deep Neural Networks (DNNs), inspired from this hierarchy of neurons and synapses in the human brain, have achieved outstanding classification accuracies across myriad cognitive applications including computer vision \cite{krizhevsky2012imagenet}, speech recognition and natural language processing \cite{mikolov2010recurrent}. Eventually, this has led to its ubiquitous presence in recognition applications across a wide variety of computational platforms that we use in our day-to-day life. For example, Siri, Google Now, Cortana are intelligent personal assistants developed by Apple, Google and Microsoft respectively and run DNNs to recognize external inputs (image, voice etc.).

The tremendous improvement in DNN performance has been possible due to the surge in the scale of the DNNs. LeCun et al. classified handwritten digits in 1998 with less than 1M synapses \cite{lecun1998gradient}. Krizhevsky et al. proposed AlexNet and won the ImageNet challenge in 2012 by using 650k neurons and 60M synapses \cite{krizhevsky2012imagenet}. In 2014, Simoyan et al. developed VGGNet using 138M synapses \cite{simonyan2014very}. Karpathy et al. proposed a DNN to convert image to natural language employing 130M Convolutional Neural Network (CNN) synapses and 100M Recurrent Neural Network (RNN) synapses \cite{karpathy2015deep}.

DNNs have powerful cognitive abilities but involve data-intensive computations leading to high power and memory bandwidth requirements. Larger DNNs require larger memory size to fit the model (weights) as well as large number of data movements between memory and the computation core. Consequently, their energy and resource requirements impede their deployment in low-power and resource constrained platforms. Furthermore, their energy consumption is orders of magnitude greater than the human brain. For example, AlexNet thrives on 2-4 GOPS (60M synaptic weights) of compute power (memory storage) per inference. Eventually, these power and memory bottlenecks posed by DNN acceleration on von-Neumann machines motivated the research on Neuromorphic Computing (NC) systems.

NC systems are based on device and circuit realizations intended to mimic the functionality of neurons and synapses in the brain. However, their CMOS implementations have been shown to suffer from inefficiencies that stem from inherent mismatch between the NC building blocks (synapse and neuron) and CMOS primitives (instructions, Boolean logic). Consequently, the CMOS realization of synapse functionality requires dozens of transistors to mimic a single synapse \cite{akopyan2015truenorth}. To this effect, device, circuit and architecture level neuromorphic designs have been explored using emerging memristive technologies \cite{jo2010nanoscale,  sengupta2016proposal, ankit2017resparc}. Memristors are programmable resistors and can encode the synaptic weights of the DNN. An MCA is a crossbar with memristive devices at its cross-points, which receives voltage inputs (at its rows) and produces an output current (at its columns) that is the weighted summation of the encoded weights at that column and the input voltage. This is a direct consequence of the Kirchhoff's law, as the current output along a column from any cross-point will be the product of the conductance at that cross-point and the voltage across it. Thus, MCA is an analog computation unit and performs highly area and energy efficient inner-product operations \cite{prezioso2015training}. The MCA outputs are interfaced with neurons to produce the neuron output. Additionally, an MCA stores the weights thereby enabling in-memory computing which enhances the energy-efficiency by circumventing the energy-hungry data movements between the memory and the computation core. Consequently, MCAs have been aggressively harnessed for energy-efficient DNN acceleration \cite{chi2016prime, shafiee2016isaac, ankit2017resparc}.

Despite the success in efficient NC system designs, the increasing scale of DNNs impede their utility in resource-constrained and low power systems. This is because the DNN topology (number of layers and neurons in each layer) is fixed before the training process starts, thereby removing the avenues of network structure optimization. Previous research has shown network pruning as an efficient technique for dynamically learning a DNN's structure (connections) during the training process \cite{han2015learning}. This produces highly sparse DNN structures, which significantly reduce the storage (memory), and computations required by the DNN, thereby making them suitable for lower power realizations. However, such algorithmic approaches are not coherent with memristive technology due to the rigid structure of an MCA. MCAs physically map the synapses (connections) in the DNN as well as act as the computation units. Pruning a synapse mapped onto an MCA merely makes the cross-point unused. This is because each cross-point physically serves as a possible connection channel between a specific input neuron (mapped onto the row) and an output neuron (mapped onto the column). Hence, despite being energy-efficient inner-product engines, the structural rigidity of an MCA degrades its utility for accelerating sparser DNNs. Consequently, realizations of such sparse DNNs onto MCA based architectures would result in highly area-inefficient designs. Typical MCA based architectures use peripherals namely buffers, communication and control logic to integrate the MCAs in order to facilitate acceleration of DNNs with varying topologies. Increasing the sparsity deteriorates the crossbar utilization, which results in a peripheral dominated energy profile. Eventually, network sparsity does not translate to commensurate area and energy savings in an MCA based architecture. In this work, we propose TraNNsformer - an integrated training framework for dynamically learning clustered connections during training. This is motivated by the observation that pruning a DNN at the MCA granularity, instead of a synapse granularity, can preserve the benefits of algorithmic pruning and network sparsity at the hardware level, in MCA based neuromorphic systems. Our approach efficiently prunes the DNN by dynamically making pruning decisions from accuracy perspective (removing unnecessary connections to maximize sparsity) as well as clustering perspective (removing unclustered synapses to maximize MCA utilization) thereby producing trained DNNs that can efficiently utilize the benefits of post-CMOS memristive technology. Further, our proposed transformation is technology agnostic as it enables mapping of connectivity structures using any MCA size permitted by the memristive technology for reliable operations.

This work focuses on efficient transformations for Fully Connected (FC) layers. Majority of the image processing and computer vision applications run CNNs, which are comprised of several convolution layers followed by a few FC layers. However, more than 96\% of the synapses are in the FC layers \cite{han2016eie}. Convolution layers are inherently sparse as each output neuron receives inputs from a fixed receptive field (not all the input neurons) equal to the kernel size. Further, this sparsity has a definite structure as these connections are known beforehand thereby allowing their efficient realizations on MCA structures. Additionally, data reuse (weight sharing) in convolution layers allows MCA reuse thereby allowing a sub-linear scaling of the neuromorphic system with respect to the DNN size. In contrast, the FC layers have no data reuse. Although, data batching helps to mitigate this while training, it is unsuitable for real-time applications with latency constraints \cite{han2016eie}. Further, the FC layers when subjected to pruning results in irregular removal of connections \cite{anwar2015structured} as it is primarily driven by accuracy constraints. FC layers are also widely used in different types of DNNs namely MLP, RNN, Long Short Term Memory (LSTM) \cite{sundermeyer2012lstm} and SNN \cite{diehl2015fast}. Hence, efficient transformation on FC layers to remove the unnecessary connections along with preserving a clustered connectivity structure is pivotal to efficient realizations of DNNs on MCA based architectures.

\begin{figure}[!t]
\centering
\includegraphics[width=3.2 in]{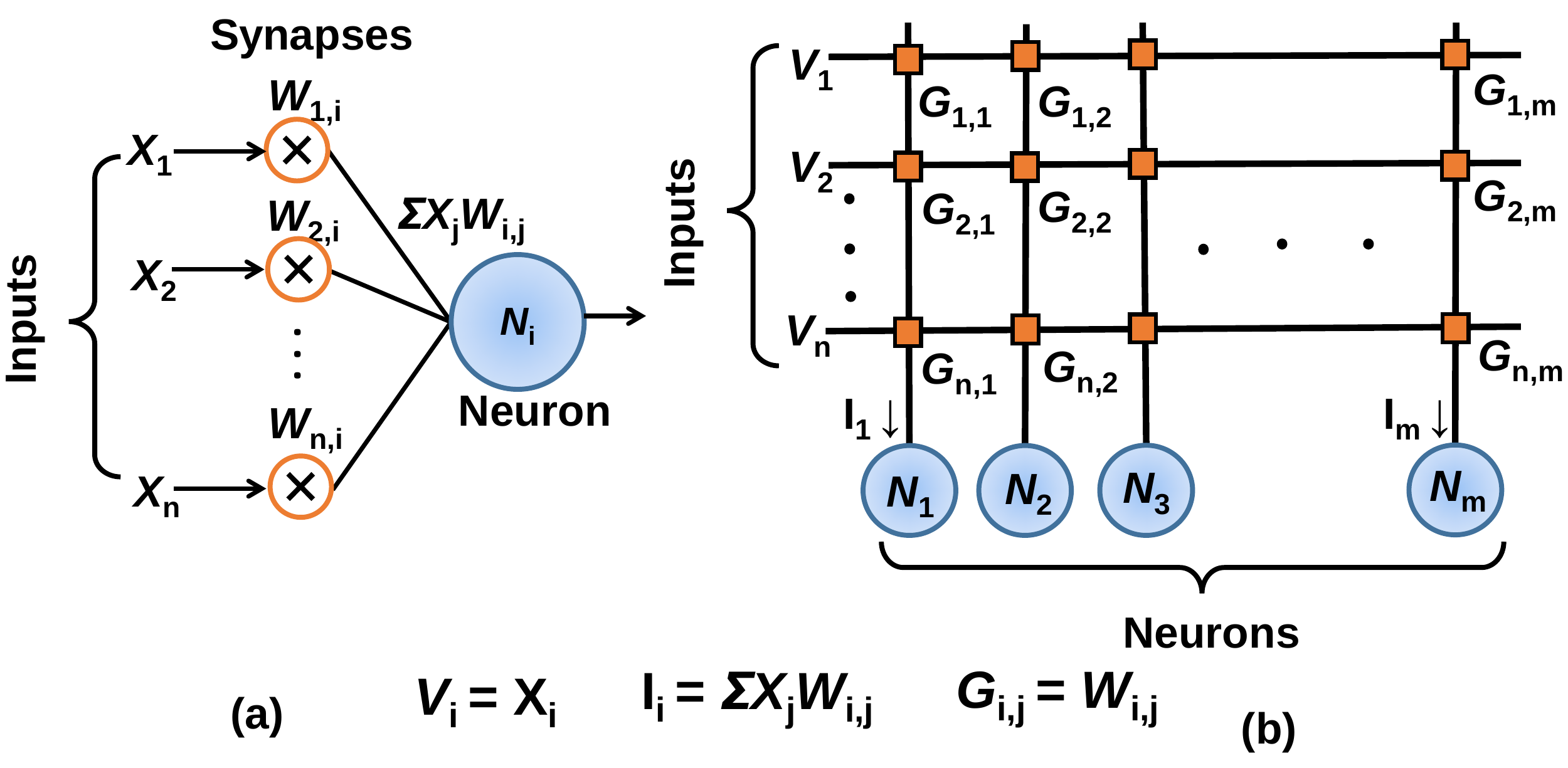}
\caption{\textbf{(a) A two-layered MLP based Neural Network (b) Neural Network mapped to Memristive Crossbar Array (MCA)}}
\end{figure}
\setlength{\textfloatsep}{5pt minus 2.0pt}

\begin{figure*}[!t]
\centering
\includegraphics[width=\textwidth]{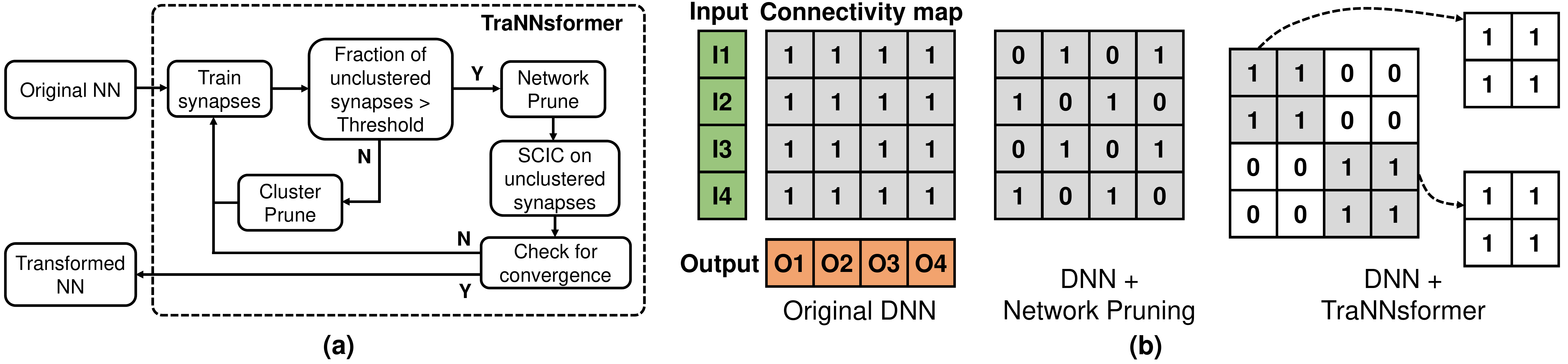}
\caption{\textbf{(a) Logical Flow Diagram of TraNNsformer Framework. The original DNN architecture during training undergoes clustering to form regions that can be mapped onto MCAs with high utilization factors, while pruning the connections that don't contribute to cluster formation. (b) Toy example to illustrate the impact of Network pruning and TraNNsformer on a DNN connectivity matrix. Network pruning leads to irregular sparsity that cannot be mapped directly onto MCAs. TraNNsformer forms smaller clusters that can be mapped onto MCAs. Note that 1/0 only represents a connection being present and not the actual value of the weight.}}
\end{figure*}

Prior work on learning structured connectivity in DNNs during the training process has focused on transformations for CMOS based architectures (Graphics Processing Unit) \cite{anwar2015structured, wen2016learning}. While, post-CMOS based MCA architectures were explored in \cite{wen2015eda}, they focused on clustering the synapses after the training process finishes. Our work distinguishes from the prior works as it proposes an integrated training framework to maximize the benefits from post-CMOS MCA based systems by designing DNN transformations, which conform to MCA's structural rigidity. Additionally, we also show that offline clustering (clustering after training) is unable to preserve the benefits of sparsity at the hardware level.

In summary, this work makes the following contributions:
\begin{enumerate} 
\item \textbf{We present a Size-Constrained Iterative Clustering} (SCIC) algorithm to enable efficient mapping of FC based connectivity matrices on MCA sizes permissible by the underlying technology.
\item \textbf{We propose TraNNsformer - an Integrated Training Framework} harnessing the SCIC algorithm to enable energy and area efficient design of DNNs for MCA based architectures.
\item \textbf{We evaluate the proposed methodology} on a wide range of benchmarks namely Digit Recognition, House Number Recognition and Object Classification using different MLP based SNNs.
\item \textbf{We analyze} the resulting energy, area benefits of TraNNsformer for post-CMOS crossbar based architectures and CMOS based general-purpose architectures.
\end{enumerate}

\section{Neural Network Basics}
Neural Networks are a class of machine learning algorithms that are comprised of multiple layers of neurons (activations) interconnected with synapses (weights). MLPs are a class of neural networks with fully connected topology i.e. each neuron in a layer receives inputs from all the neurons in the previous layer. A neuron receives inputs that are modulated based on the synapse connecting the input and output neuron. Subsequently, it performs a non-linear computation on the received inputs to produce the output, which is sent to the neurons in the successive layers. Fig. 1(a) shows a two-layered MLP topology being mapped onto an MCA (shown in Fig. 1(b)).

SNN is regarded as the third generation neural network. SNNs require the input to be encoded as spike trains and involve spike-based (0/1) information transfer between neurons. At a particular instant, each spike is propagated through the layers of the network while the neurons accumulate the spikes over time causing the neuron to fire or spike.

\section{TraNNsformer Framework}
In this section, we discuss in detail about the TraNNsformer framework (shown in Fig. 2(a)), its effect on DNN sparsity (shown in Fig. 2(b)) and the resulting benefits on two types of architectures 1. MCA based architecture and 2. CMOS based general-purpose architecture. Subsection 3.1 describes the Size Constrained Clustering Algorithm (SCIC), which converts a DNN's connectivity structure into a set of high utilization clusters that can be mapped onto MCAs. Subsection 3.2 details TraNNsformer, which is the Integrated Training approach, built on SCIC to transform the DNN connectivity matrix into an optimally sparse and maximally clustered structure during the training process. Subsection 3.3 discusses the benefits of TraNNsformer on area and energy consumption for MCA based architectures. Subsection 3.4 discusses the impact of TraNNsformer for CMOS based general-purpose architectures.

\subsection{Size Constrained Iterative Clustering (SCIC)}
Size constrained Iterative Clustering is an adaptation of the Spectral Clustering algorithm \cite{ng2001spectral} that generates clusters from a connectivity matrix. Connectivity matrix (C) is a (0, 1)-matrix that represents the morphology of a layer in the DNN such that a value C\textsubscript{ij} being one corresponds to a non-zero synapse between the ``i\textsubscript{th}'' input neuron and ``j\textsubscript{th}'' output neuron. Utilization factor is the fraction of used (mapped) cross-points in an MCA. A zero value in a cluster would result in an unused cross-point (if the cluster is mapped to MCA).

\begin{algorithm}[h]
\small
\caption{Spectral Clustering (SC)}
Input: Similarity matrix \textit{S $\in$ R\textsuperscript{n$\times$n}} for the graph (a row in \textit{S} corresponds to a graph node), \textit{K} clusters to construct
\begin{algorithmic}[1]
\State $ \text{Compute the degree matrix: } \textit{D}$
\State $ \text{Compute the normaized laplacian matrix: } \textit{L}$
\State $ \text{Perform an eigenvalue decomposition: } \text{$\mathit{L = U \Sigma U\textsuperscript{T}}$ }$
\State \parbox[t]{\dimexpr\linewidth-\algorithmicindent}{Extract the \textit{K} columns of \textit{U} corresponding to the \textit{K} 
        \text{smallest eigenvalues to form:} \text{$\widetilde{U}$}}
\State $ \text{Cluster the row vectors of $ \widetilde{U}$ using } \textit{K-means } \text{algorithm}$
\end{algorithmic}
Output: \textit{K} clusters - a row vector of $\widetilde{U}$ corresponds to a row of \textit{S}
\end{algorithm}

Spectral Clustering (SC) is a graph clustering algorithm that produces a set of disjoint graph nodes such that intra (inter) cluster associativity is maximized (minimized). We adopt spectral clustering to cluster the connectivity matrix for a DNN layer where the input and output neurons are the graph nodes and a synapse corresponds to a graph edge. As shown in Algorithm 1, a symmetric matrix (L) is defined on the graph (connectivity matrix) using the adjacency matrix (S) and degree matrix (D). Subsequently, it undergoes an eigenvalue decomposition followed by dimensionality reduction.  Finally, the remaining row vectors are clustered using the K-means algorithm. The intra-associativity maximization produces clusters that can be mapped to MCAs with high utilization factor. Furthermore, inter-associativity minimization enables low overhead or high throughput (number of neuron outputs computed per cycle) integration of MCA currents for generating a neuron output. This is because the synapses corresponding to a particular output neuron can be spread across multiple clusters (mapped across multiple MCAs). Eventually, the crossbar output currents from multiple MCAs are integrated onto the output neuron in a time-multiplexed fashion (shown in Fig. 3) to generate the final neuron output. Hence, minimizing the inter-cluster associativity results in a commensurate decrease in the inter MCA interaction thereby maximizing the throughput.

\begin{figure}[!t]
\centering
\includegraphics[width=3 in]{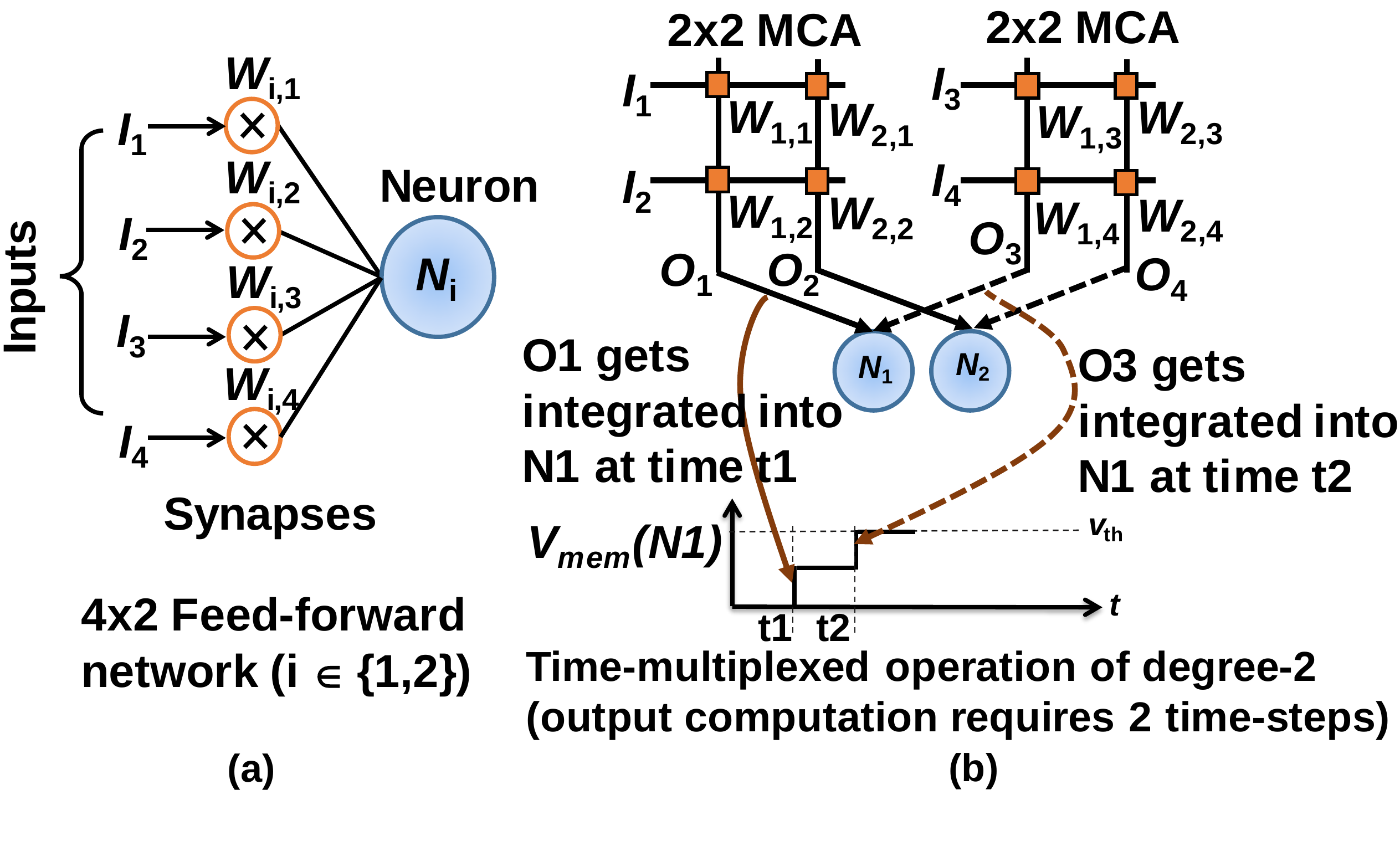}
\caption{\textbf{(a) A feed-forward neural network with neuron fan-in of 4 (b) Mapping the 4 fan-in neurons using a 2$\times$2 MCAs}}
\label{svhn_unclus}
\end{figure}

\begin{algorithm}[!b]
\small
\caption{Size Constrained Iterative Clustering (SCIC)}
Input: Connectivity matrix \textit{C $\in$ R\textsuperscript{m$\times$n}} (\textit{m} and \textit{n} are number of input and output neurons respectively), crossbar\_size, base\_util\_factor, min\_util\_factor 
\begin{algorithmic}[1]
\State $ \text{Initizalize: cluster\_set = \textit{C} , num\_clusters = \textit{k}}$
\While {$\Delta$cluster\_set $\ne$ 0}
	\For{all \textit{cluster} in \textit{cluster\_set}}
		\If{sizeof (\textit{cluster}) $>$ crossbar\_size}
  			\State $\text{clusters\_temp $\gets$ $\emptyset$}$
			\State $\text{util\_factor = base\_util\_factor}$
			\While {clusters\_temp $\equiv$ $\emptyset$}
				\If {util\_factor $<$ min\_util\_factor}
					\State $\text{break}$
				\Else
					\State $\text{decay (util\_factor)}$
					\State $\text{$\widetilde{C}$ = connectivity matrix of cluster}$ 
		    			\State $\text{construct similarity matrix \textit{S} from $\widetilde{C}$}$
					\State $\text{clusters\_temp = spectral\_clustering(\textit{S}, \textit{k})}$
					\State $\text{clusters\_set $\gets$ cluster\_set + clusters\_temp}$
					\State $\text{update \textit{C} to reflect unclustered synapses}$
				\EndIf
			\EndWhile
		\EndIf
      	\EndFor
\EndWhile
\end{algorithmic}
\hspace{-4mm} Output: \textit{cluster\_set}
\end{algorithm}

SCIC is an iterative algorithm that minimizes the number of unclustered synapses while ensuring cluster generation with higher utilization factors. As shown in Algorithm 2, each iteration of SCIC algorithm runs SC on the remaining connectivity matrix until the reduction in unclustered synapses ceases. Further, in each iteration, we greedily select high quality clusters (clusters that map to MCAs with high utilization factor) that meet a specific quality threshold. Other clusters are ignored and merged with the existing connectivity matrix to explore new clustering avenues in the subsequent iterations. The threshold subsequently decays when the formation of new clusters slows down. Hence, the greedy and iterative approaches synergistically ensure efficient clustering to produce high quality clusters. As mentioned before, MCA sizes are limited by the memristive technology to ensure their reliable operation. SCIC iteratively breaks down (forms sub clusters) from the existing clusters until they meet the MCA size specified by the given technology. Thus, SCIC enables a technology-aware mapping of connectivity matrices onto MCAs.

\subsection{Integrated Training Approach}
As shown in Algorithm 3, each training iteration (back-propagation) is accompanied with pruning followed by SCIC. Pruning removes the synapses that do not affect the accuracy of the DLN. The result of pruning is encoded as a ``prune map'' which is a (0, 1)-matrix (similar to the connectivity matrix) where ``0s'' represent a pruned synapse. ``0s'' in the prune map correspond to Accuracy Don't Cares (ADCs). Subsequently, clusters produced in SCIC are used to form a ``cluster map'' that denotes if a synapse in the connectivity matrix is part of a previously formed cluster. ``0'' in the cluster map represent Clustering Don't Cares (CDCs). The union of cluster map and prune map is masked from being pruned in the subsequent iterations.  The synapses, which belong to both the ADC and CDC set, are aggressively pruned to maximize pruning without affecting the cluster quality. The subsequent training iteration tries to recover the accuracy loss incurred due to pruning.

Although SCIC generates high quality clusters, it leaves a large fraction of synapses unclustered. This results in large number of MCAs with low utilization factors being mapped to the unclustered synapses, thereby diminishing the benefits of SCIC. Consequently, a training algorithm based on offline clustering i.e. clustering the synapses at the end of the training process will suffer from inefficiencies resulting from higher fraction of unclustered synapses. However, it is interesting to note that subsequent pruning of the synapses belonging to the intersection of ADC and CDC set removes several unclustered synapses. Furthermore, this pruning exposes new avenues for SCIC to generate high quality clusters from the remaining unclustered synapses.  TraNNsformer is inspired from this observation to synergistically combine the benefits of network pruning and SCIC to make the DNNs as sparse as possible while ensuring the clustered structure to produce MCA mappings with high utilization factors. Thus, TraNNsformer allows to dynamically learn the DLN structure in a clustered way during the training process in order to produce an optimized network for MCA based architectures.

It is worth noting that TraNNsformer favors cluster formation in the beginning to minimize the fraction of unclustered synapses. Once the unclustered synapses have been significantly reduced, TraNNsformer initiates cluster pruning (as shown in Algorithm 3). Cluster pruning incrementally prunes an entire cluster based on a combined score that quantifies cluster quality and the cluster's contribution to output accuracy. Although, cluster pruning may lead to accuracy degradation, subsequent training iteration ensures a graceful recovery of the lost accuracy. It is worth noting that cluster pruning does not affect the overall MCA utilization as it entirely removes the mapped MCA. Thus, cluster pruning allows achieving higher network sparsity (similar to network pruning) while ensuring the clustered structure of the DLN's connectivity matrix.

\begin{algorithm}[h]
\small
\caption{TraNNsformer}
\begin{algorithmic}[1]
\State $\text{Train the connectivity for an epoch}$
\If{\textit{num\_unclustered\_synapses $<$ threshold}}
	\If{\textit{training\_error $<$ training\_error\_previous}}
		\State $\text{cluster\_prune()}$
	\EndIf
\Else
	\If{\textit{training\_error $<$ training\_error\_previous}}
		\State $ \text{Prune and update \textit{prune map}}$
	\EndIf
	\State $ \text{Run SCIC on unclustered synpases and update \textit{cluster map}}$
\EndIf
\State $ \text{connectivity $\gets$ (prune map $\cup$ cluster map)}$
\State $\text{Go to 1 if convergence is not reached}$
\end{algorithmic}
\end{algorithm}

\subsection{Impact on Crossbar based architecture}
Crossbar-based architectures for DNN acceleration are comprised of computation cores each of which consists of MCAs and peripherals associated with an MCA namely buffers, communication and control logic \cite{ankit2017resparc}. Hence, the number of cores (\textit{num\_core}) has a linear dependence on the number of MCAs (\textit{num\_mca}) as shown in eqn. 1 (where \textit{``k''} is a micro-architecture dependent constant). TraNNsformer enables technology aware optimization to learn an optimally clustered network structure such that a learnt cluster can be mapped onto an MCA with high utilization factor. Consequently, it ensures that the network sparsity efficiently translates to reduction in the number of MCAs required to map the transformed DNN with respect to the original DNN. This results in a commensurate reduction in the number of cores thereby leading to area savings. 
\begin{equation}num\_core = \dfrac {num\_mca} {k} \end{equation}
The energy profile for an MCA based architecture is comprised of MCA energy (includes neuron energy) and peripheral energy components. Hence, the total energy consumption for a single inference of DNN execution can be defined as shown in eqn. 2.
\begin{equation}\small\textit{Total Energy} = \sum_{i=0}^{\text{all cores}} \textit{MCA Energy + Peripheral Energy}\end{equation}
An increase in DNN sparsity results in a corresponding decrease in the total MCA energy component irrespective of the connectivity structure. However, the total peripheral energy component depends on the number of MCA (\textit{num\_mca}) being used. As discussed before, network pruning does not lead to significant reductions in the number of MCAs due to the irregular nature of sparsity pattern, thereby not affecting the total peripheral energy. This reduces the overall energy benefits that can be obtained by highly sparse DNN connectivity structures. Additionally, the total energy profile becomes peripheral energy dominated, which would prevent harnessing the energy benefits from further efficient memristive technologies. On the contrary, TraNNsformer helps to obtain commensurate reductions in MCA energy as well as the peripheral energy component, which in turn results in significant savings in total energy consumption. Consequently, the energy profile shows a favorable distribution between the MCA and peripheral energy components.

\begin{figure}[!b]
\centering
\includegraphics[width=3.6 in]{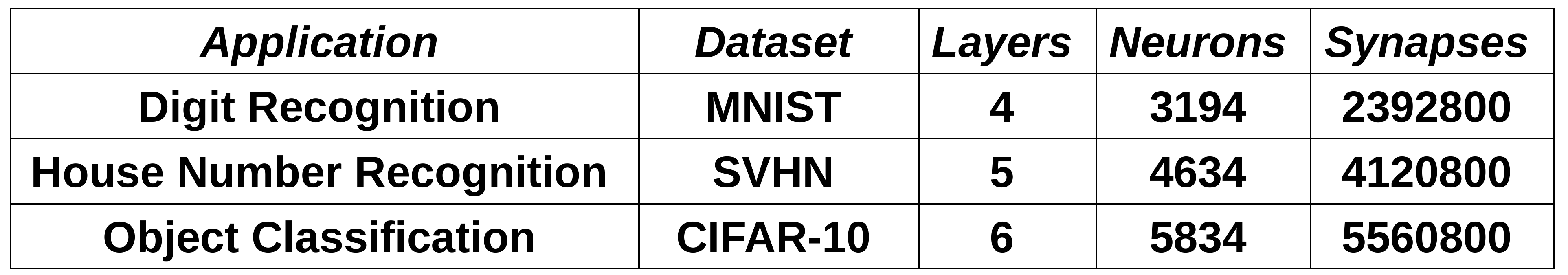}
\caption{\textbf{MLP based SNN benchmarks}}
\end{figure}
\setlength{\textfloatsep}{5pt minus 2.0pt}

\begin{figure*}[!t]
\centering
\includegraphics[width=\textwidth]{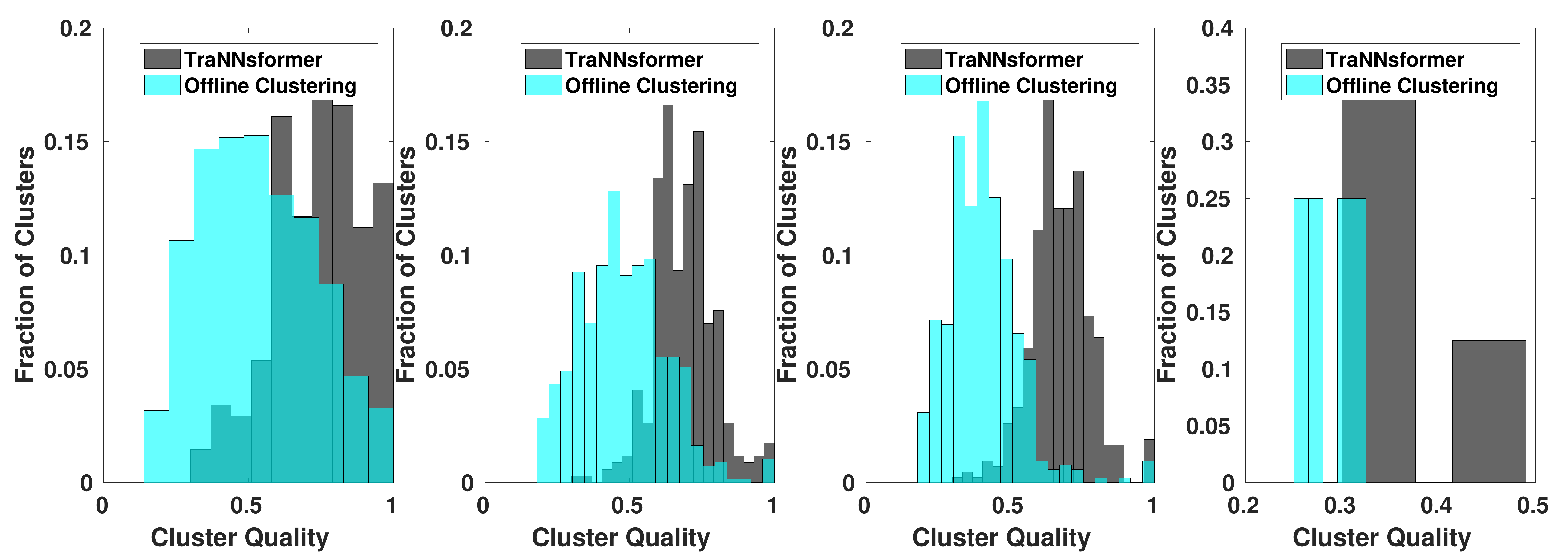}
\caption{\textbf{Comparison of crossbar utilization (cluster quality) between Offline Clustering and TraNNsformer}}
\end{figure*}
\setlength{\textfloatsep}{5pt minus 4.0pt}

\subsection{Impact on CMOS based general-purpose architecture}
Typical digital CMOS based general-purpose architectures for DNN execution consist of a memory unit to store the weights (synapses) and a computation core to perform neuronal computations using the weights fetched from the memory \cite{han2016eie}. The total energy consumption per DNN inference in such cases can be described as eqn. 3, where \textit{``n''} is the number of synapses, \textit{Computation} corresponds to the neuron energy expended per computation (multiplication and accumulation), \textit{Memory Access} corresponds to the energy spent for fetching a weight from memory and \textit{Leakage} represents the overall leakage energy consumed per inference.
\begin{equation}\small\textit{Energy} = \sum_{i=0}^{\text{n}} \textit{(Computation + Memory Access)} \textit{ + Leakage}\end{equation}
Network pruning reduces the number of synapses (\textit{``n''}) thereby reducing the number of computations and memory accesses required for DNN computation. However, this does not lead to significant reduction in the memory size, as the zero weights are an instrumental part of the DNN topology information. Consequently, DNN sparsity does not lead to significant energy savings as both memory access and leakage energies are strong functions of memory size. Additionally, energy profiles of FC layers are memory dominated which further reduces the overall energy savings. In contrast to network pruning, TraNNsformer helps to reduce the number of synapses as well as the memory storage requirements for remaining synapses. As shown in Fig. 2(b), such a transformed DNN execution resembles execution of several clusters, where, each cluster's execution is identical to a smaller DNN structure itself (except for the final activation computation). Eventually, the partial products generated from these smaller DNNs will need to be synchronized (combined) together to complete an inference of transformed DNN execution. Although, this synchronization is associated with an overhead, the overall saving from smaller memory size outweighs the overhead. Thus, TraNNsformer achieves better energy efficiency compared to the original as well as pruned DNN connectivities on CMOS based general-purpose architectures. Kindly note that other techniques based on network compression have also been studied to remove zero weight storage and optimize the memory size requirements. However, significant modifications are required at the hardware level to accelerate such compressed DNNs \cite{han2016eie}. On the contrary, TraNNsformer helps to utilize the benefits from a reduced memory size on the original accelerator (uncompressed DNNs) itself.

\section{Experimental Methodology}
TraNNsformer framework was designed using MATLAB by utilizing the relevant components from MATLAB DeepLearn Toolbox \cite{palm2012prediction}. We analyzed the algorithmic benefits on SNNs with MLP (fully-connected) topologies. To analyze the scalability of the proposed framework, we used a range of applications with different complexity namely Digit Recognition (MNIST dataset \cite{lecun1998gradient}), House Number Recognition (SVHN dataset \cite{netzer2011reading}) and Object Classification (CIFAR10 dataset \cite{krizhevsky2009learning}). Different network architectures commensurate to the dataset complexity were chosen for each application (shown in Fig. 4) to achieve high classification accuracy. Although we analyze our results for SNN based benchmarks, the algorithmic benefits would be similar for Artificial Neural Networks (ANNs). This is because TraNNsformer works on optimizing the connectivity structure between layers in a DNN, which is similar for both SNN and ANN. Kindly note that ANN and SNN (used in our case) differ only in the way inputs are transmitted between layers.

We used the architecture proposed in \cite{ankit2017resparc} for studying the system-level benefits of TraNNsformer on post-CMOS MCA based architectures. For the memristive devices, we used a resistance range of ``20K$\Omega$ -- 200K$\Omega$'' with 16 levels (4 bits) for weight-discretization, that is typical of memristive technologies such as Phase Change Memory (PCM), Ag-Si \cite{rajendran2013specifications}. We considered an operating voltage of ``Vdd/2'' for the MCA as it is interfaced with CMOS neurons \cite{joubert2012hardware}. To analyze the system level benefits on CMOS based general-purpose architectures, we use the energy numbers for arithmetic operations in a 45nm CMOS process shown in \cite{han2015learning}. The memory for weight storage was modeled using CACTI \cite{muralimanohar2007optimizing}.

TraNNsformer is targeted to improve the area and energy consumption of DNNs during inference phase but has higher training effort (in terms of time and energy consumption) than network pruning. However, typical DNNs are trained very infrequently but used for testing/inference for much longer times.

\vspace{-4mm}
\section{Results}
In this section, we present the results of various experiments that demonstrate the benefits of TraNNsformer (at algorithm and system level) for DNN acceleration on post-CMOS MCA based systems. Note that we have used normalized values to report the area and energy benefits. This is because the benefits from our proposed framework are orthogonal to the benefits obtained from any particular choice/design of MCA based architecture.

Additionally, we also evaluate the benefits obtained on CMOS based systems to demonstrate the effectiveness of the proposed framework towards translating the benefits of DNN sparsity to reduced memory storage requirements.

\subsection{Algorithm level analysis}
Fig. 5 shows the fractional distribution of clusters formed with respect to the cluster qualities for two different training approaches namely TraNNsformer and offline clustering on SVHN dataset. In both approaches, DNNs are trained to achieve iso-accuracies. Cluster quality represents the number of non-zero synapses present in a cluster. Hence, a cluster with higher cluster quality will map to an MCA with high utilization factor. As mentioned before, higher MCA utilizations help to achieve higher area as well as energy efficiency on MCA based architectures. Fig. 5 represents the MCA utilization for a DNN with 70\% average sparsity (across all layers) for offline clustering. MCA utilization for a DNN trained with network pruning (pruned DNN) is almost uniform across all MCAs that are required for mapping. Consequently, the pruned DNN maps to MCAs with 0.3 utilization factor. As shown in Fig. 5, the pruned DNN upon undergoing offline clustering significantly improves the MCA utilization across all the layers. It can also be seen that, TraNNsformer further improves the MCA utilization by a significant factor across all layers in comparison to the offline clustering approach. This underscores the fact that dynamic cluster formation during the training process helps to get highly structured DNN connectivities in comparison to both pruning and offline clustering.

Fig. 6 shows that the fraction of unclustered synapses remaining after the training process in offline clustering approach is significantly higher than TraNNsformer across all the layers (shown by markers in Fig. 6) of DNN. A large number of unclustered synapses is undesirable, as their mapping results in large number of poorly utilized MCAs. Additionally, the number of MCAs required for mapping unclustered synapses are much higher than the number of MCAs mapped to the clustered synapses thereby, diminishing the overall benefits obtained from clustering. However, TraNNsformer based DNN training leads to much smaller fraction of unclustered synapses across all the layers in the DNN (except the last layer). Consequently, the number of MCAs with low utilization factors are insignificant with respect to the number of MCAs that are mapped to the clustered synapses.

\begin{figure}[h]
\centering
\includegraphics[width=3 in]{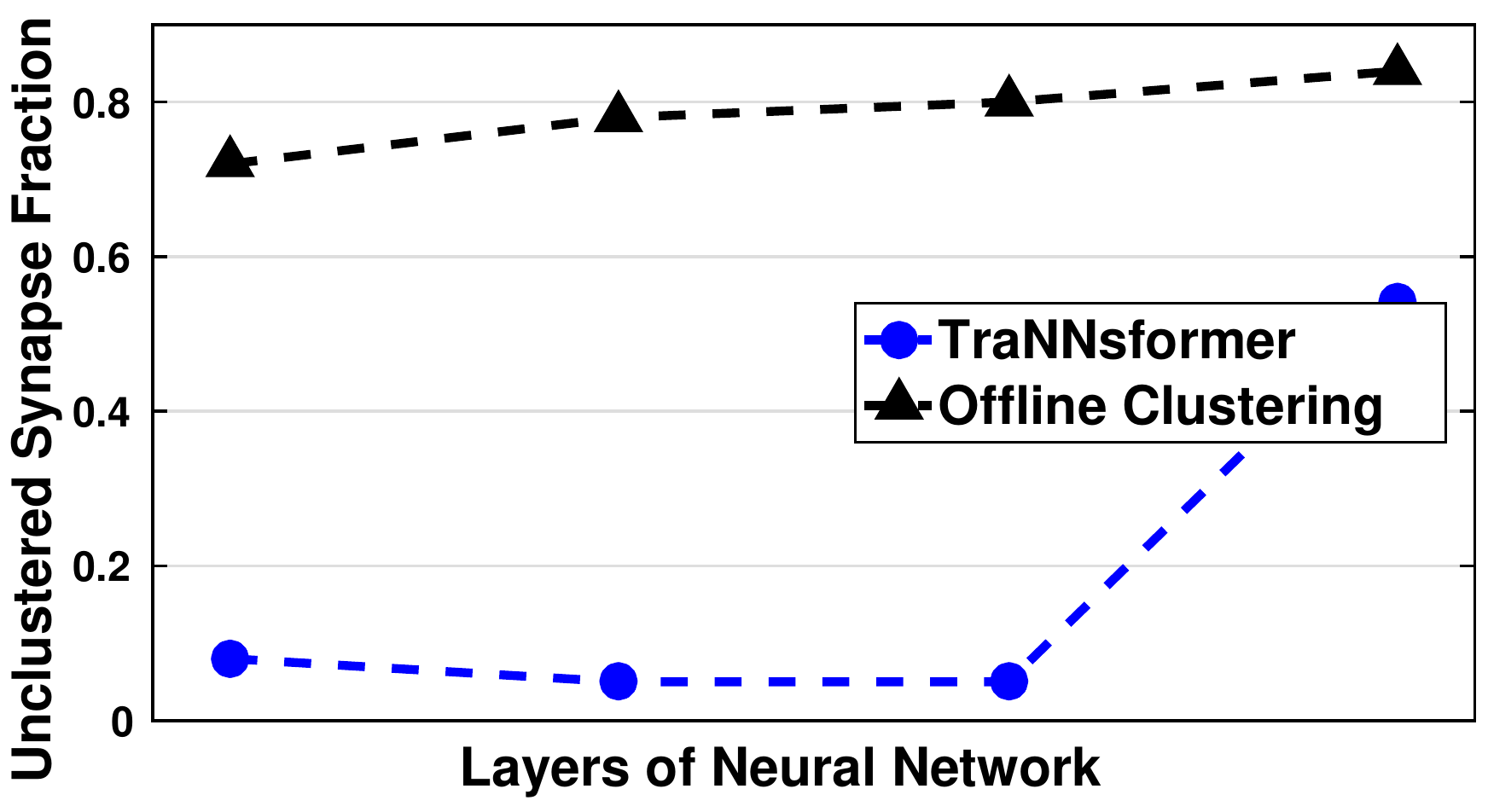}
\caption{\textbf{Comparison of fraction of unclustered synapses between Offline Clustering and TraNNsformer (the data points correspond to the layers of DNN). Note that the last fully connected layer consists of a small fraction of synapses (\textless 1\%), thereby having insignificant effect on overall unclustered synapse comparison.}}
\end{figure}

Both lower fraction of unclustered synapses and higher cluster quality are equally important factors in reducing the number of MCAs required for mapping. A DNN with low cluster quality would map the clustered synapses across a large number of MCAs whereas a DNN with higher fraction of unclustered synapses consumes a large number of MCAs to map the unclustered synapses. Thus, TraNNsformer optimizes both these factors concurrently to reduce the total number of MCAs required. Please note that similar results for cluster quality distribution and fraction of unclustered synapses were obtained for MNIST and CIFAR10 as well thereby underscoring the scalability benefits of the proposed training framework with respect to DNN size (number of layers, number of neurons in each layer).

\subsection{Area and energy comparisons on MCA based architecture}
\begin{figure}[!t]
\centering
\includegraphics[width=3.2 in]{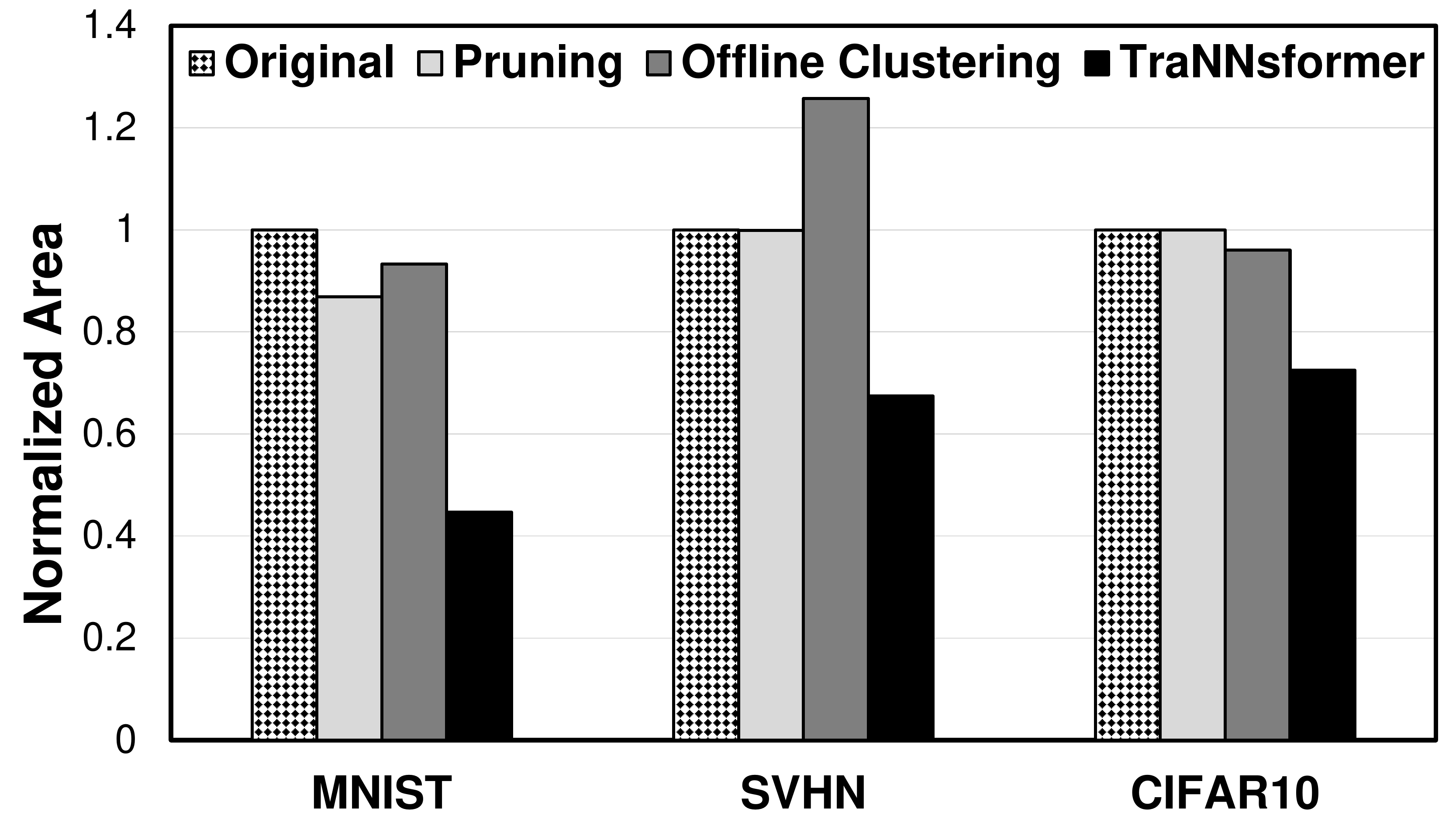}
\caption{\textbf{Comparison of area consumption on MCA based architecture for different DNN training approaches}}
\end{figure}

Fig. 7 shows the area consumption on MCA based architecture for DNNs with iso-accuracies trained using four different approaches namely 1) Original i.e. typical back-propagation 2) Pruning (network pruning) 3) Offline clustering and 4) TraNNsformer. The energy consumption has been normalized with respect to the ``Original'' area consumption for each dataset. It can be seen that TraNNsformer achieves area savings of 28\% -- 55\% (39\% on average) compared to the original DNN across all datasets. Furthermore, TraNNsformer also achieves significant area reductions of 28\% -- 49\% (37\% on average) with respect to the pruned DNNs across all datasets. This underscores the effectiveness of the proposed TraNNsformer framework in preserving the benefits of DNN sparsity at the hardware level. As mentioned before, a DNN trained with network pruning has irregular (unstructured) sparsity, which results in the DNN being mapped across a large number of MCAs with low utilization factors. Consequently, the benefits of network sparsity does not improve the DNN's area efficiency. It can be seen that the area consumption for DNNs trained with offline clustering is significantly higher than the TraNNsformer case across all benchmarks. This justifies the importance of a clustering driven DNN training approach towards achieving efficient implementation on MCA based systems. It is also worth noting that the area consumption in offline clustering case is irregular i.e. lower or comparable to network pruning in some cases (CIFAR10) while being higher in other cases (MNIST and SVHN). This is because the larger fraction of unclustered synapses remaining after clustering in SVHN gets mapped to a large number of MCAs with very utilization factors thereby, worsening the area consumption.

\begin{figure}[!b]
\centering
\includegraphics[width=3.2 in]{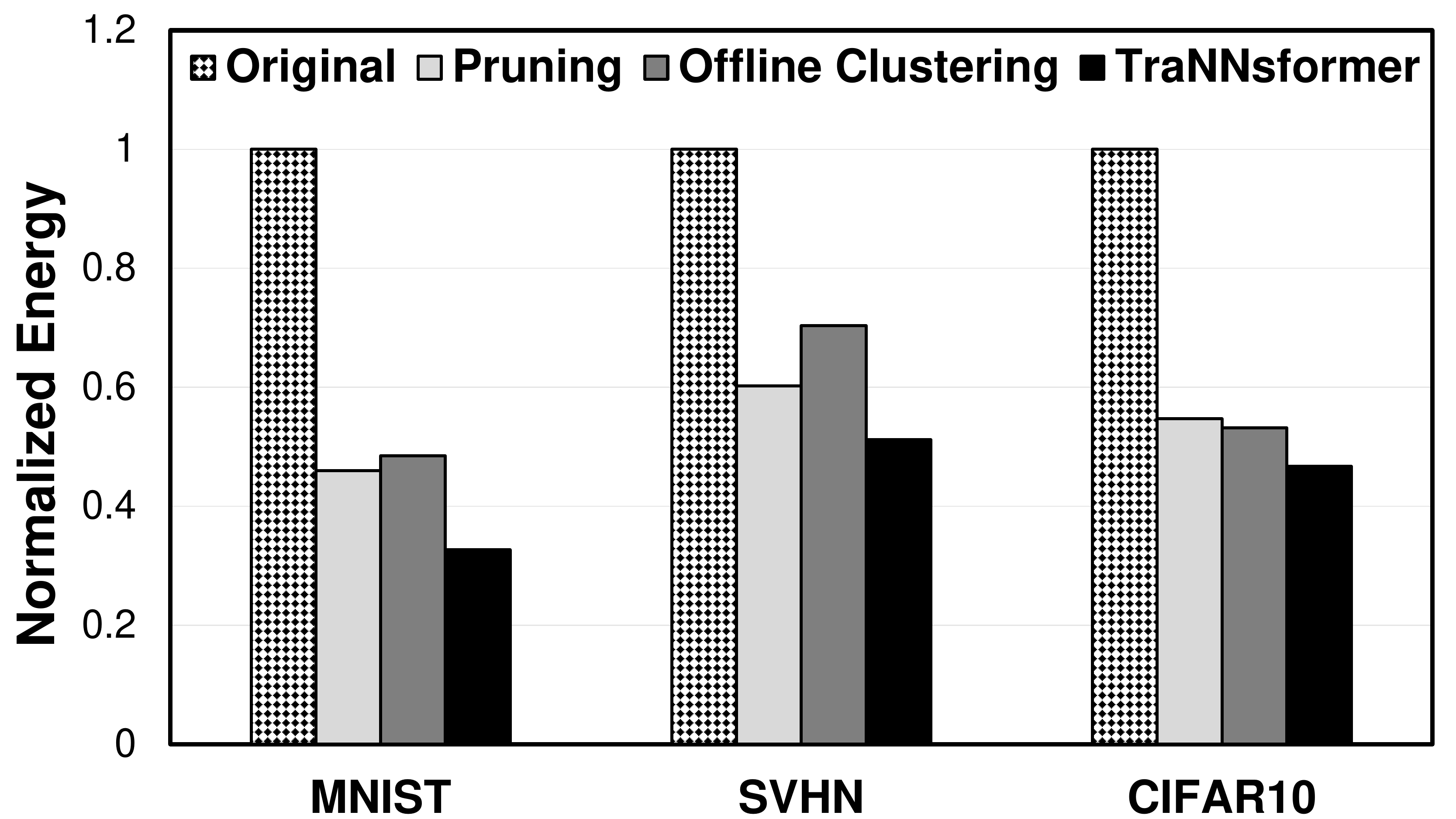}
\caption{\textbf{Comparison of energy consumption on MCA based architecture for different DNN training approaches}}
\end{figure}

Fig. 8 shows the energy consumption per classification obtained for DNNs trained using the four training approaches. The energy consumption has been normalized with respect to the ``Original'' energy consumption for each dataset. It can be seen that TraNNsformer achieves significant energy improvements of 49\% -- 67\% (56\% on average) across all datasets. Furthermore, it also achieves 15\% -- 29\% (20\% on average) energy reduction with respect to the pruned DNNs across all datasets. As mentioned before, pruning decreases the MCA energy component only while having minimal effect on the peripheral energy component. However, TraNNsformer based network sparsity translates to commensurate savings for both MCA as well as peripheral energy components thereby, leading to greater energy savings. It can also be seen that offline clustering approaches have higher energy consumption (MNIST, SVHN) than the pruned DNNs owing to the higher fraction of unclustered synapses.

\subsection{Energy comparison of CMOS based general-purpose architectures}
\begin{figure}[h]
\centering
\includegraphics[width=2.75 in]{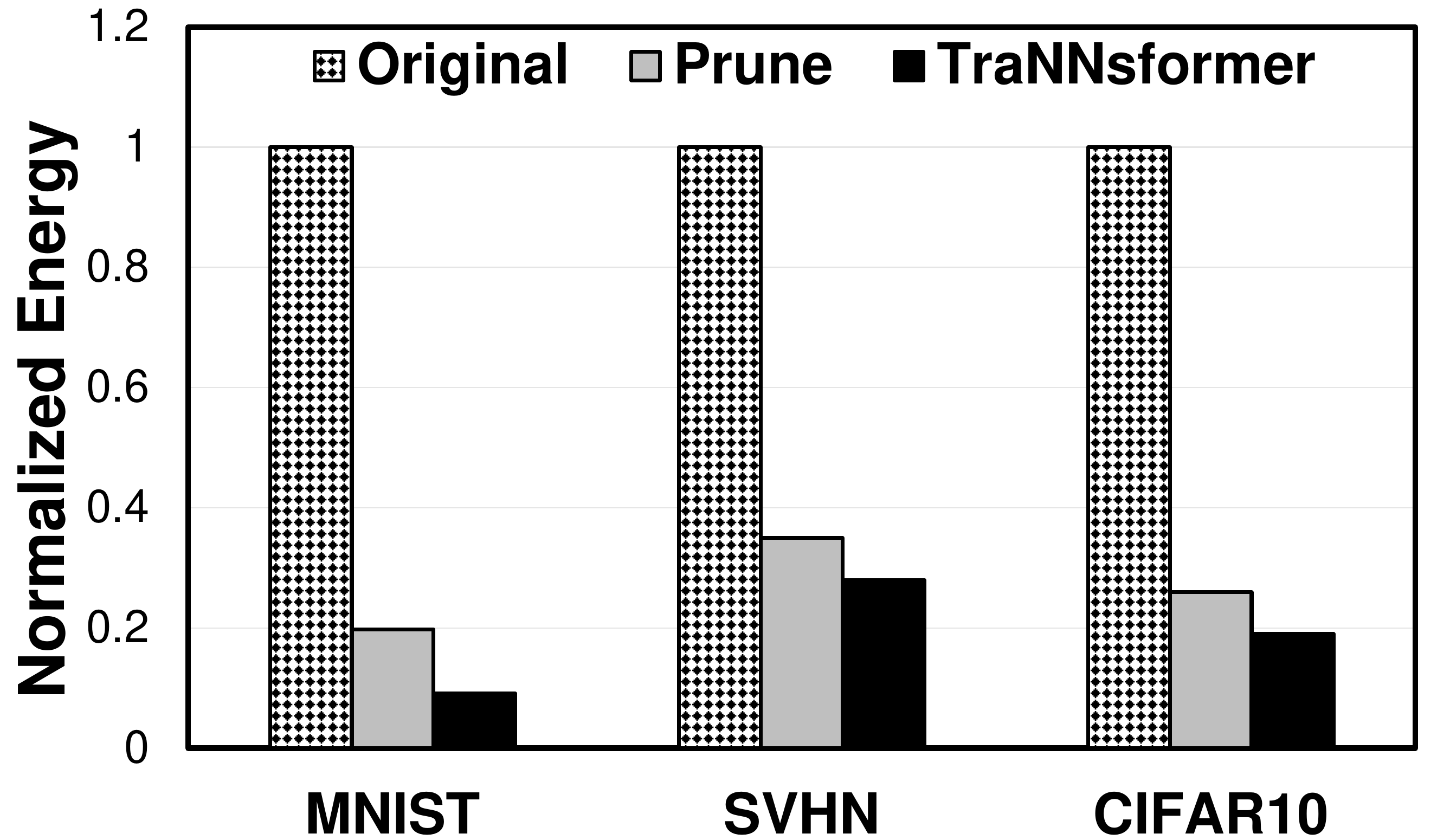}
\caption{\textbf{Comparison of energy consumption on CMOS based general-purpose architecture for different DNN training approaches}}
\end{figure}
Fig. 9 shows the energy consumption per classification of DNNs trained using three different training approaches 1) Original i.e. typical back-propagation 2) Pruning (network pruning) and 3) TraNNsformer. The energy consumption has been normalized with respect to the ``Original'' energy consumption for each dataset. It can be seen that both network pruning and TraNNsformer achieve significant energy reductions in comparison to the original DNN. However, TraNNsformer is 20\% -- 57\% (37\% on average) more energy efficient compared to the pruned DNN. This efficiency stems from the significant reduction in memory storage requirement for TraNNsformer based DNNs. A reduction in memory size translates to commensurate savings in memory energy components (both access and leakage). Additionally, as mentioned before, the energy profiles for FC layers are memory energy dominated which results in the memory energy savings being translated to significant savings in overall energy consumption.

\section{Conclusions}
The intrinsic compatibility of post-CMOS technologies with biological primitives has ushered the usage of Memristive Crossbar Arrays (MCAs) in neuromorphic systems in order to achieve low-power acceleration of Deep Neural Networks (DNNs). However, DNNs have multiple static (known before training) connectivity patterns (for instance, CNN, MLP, RNN) which are primarily application dependent. Further, techniques to optimize a DNN by obtaining highly sparse connectivity (network pruning) adds a high degree of dynamic variability (not known before training) to the connectivity pattern. This variability in connectivity pattern requires hardware-aware mapping algorithms to maximize the area and the energy benefits for MCA based systems. While rule based mapping techniques can address the static variability, dynamic connectivity patterns are much more challenging to map owing to their large degree of variability. Additionally, an inefficient mapping algorithm prevents the algorithmic benefits of sparsity to be preserved at the hardware level. In this work, we proposed TraNNsformer an integrated training framework that learns connectivity structures, which can be efficiently mapped to MCAs while preserving the algorithmic benefits of network sparsity. We also developed a technology-aware clustering approach to produce efficient mappings for any MCA size, permissible by the technology for reliable operations. Furthermore, we show that TraNNsformer also leads to energy reductions in CMOS based general-purpose architectures, thereby proving the generality of the proposed framework across different architecture styles.  Our results on a range of recognition applications suggest that TraNNsformer is a promising framework to implement DNNs, providing a scalable solution to designing large-scale neuromorphic systems.

\vspace{-3mm}
{\footnotesize
\bibliography{bibfile}}

\end{document}